# Toward a High-Performance IPMC Soft Actuator using A disturbance-aided method

Mohsen Annabestani*, Mohammad Hossein Sayad, Pouria Esmaeili-Dokht, and Mehdi Fardmanesh*

*Abstract*— Besides the advantages of Ionic polymer-metal composites (IPMCs) for biomedical applications, there are some drawbacks in their performance, which can be enhanced. One of those critical drawbacks is "back relaxation" (BR). If we apply a step voltage to IPMC, it will bend in the anode direction. Afterward, there is an unwanted and relatively slow counter-bending toward the cathode side. There are some disadvantages in the current BR control methods of IPMC actuators that prevent them from being used in real applications. This paper presents a new non-feedback method for eliminating the BR effect of non-patterned IPMCs by using a relatively high-frequency disturbance and proving it by theoretical and experimental explanations. The results show that the proposed method, needless to have any pattern on the electrodes of the IPMCs, can significantly eliminate the BR effect. Unlike the patterned IPMCs, no reduction will occur in the bending amplitude of IPMC, and even we can see the increased bending amplitude.

## I. INTRODUCTION

IPMCs are electroactive polymers used in various applications, especially biological applications. Because of its promising features and electrically controllability, it has a wide range of applications [1-8]. IPMC has some drawbacks which should be considered in its applications. The most significant disadvantage of this soft material is the back relaxation phenomenon. When a step voltage is applied to a Nafion-based IPMC, it will bend toward the anode. Also, there is an undesirable and slow counter bending toward the cathode simultaneously. This unwanted bending is called Back Relaxation (BR) effect [9, 10]. To clear out this phenomenon, considering applying a step voltage to IPMC. IPMC should bend toward the anode and remains steady at the final bending position if everything was ideal. However, because of the BR effect, IPMC did not reach the final position and, before there, start to bend back to its starting position. This bending continues, and IPMC stops from bending and becomes stable somewhere between the final and initial positions [11]. In one of our previous works [11], we proposed a way to restrain the BR effect using patterned electrode IPMC and applied Gaussian disturbance as the control signal. There are also some examples for fabrication-based methods, from consuming Flemion polyelectrolyte instead of Nafion for IPMC's membrane [12] to adding a buffer layer between platinum and Nafion, made of Palladium [13]. Another example was an proper mixture of the solvent and cations [14]. The second category contains the control-based method, including a model-based closed-loop precision position control [15], using two(or more) IPMCs for a control-based method [16], and using control units for patterned IPMCs such as feedforward and feedback controllers [17]. Each of these introduced methods has at least one constraint, which limited the practical application of the proposed method. It was shown in Nemat-Nasser's work [4] that there is no BR effect in Flemion-based IPMCs, which was observed on Nafion-based IPMCs. Also, in Flemion-based IPMCs, bending is continuous. Still, there is more commercial availability and more practical advantages of Nafion over Flemion. We should consider that continuous bending is not always helpful. As another example, in [17], it was shown that the feedback acquisition method is applicable only for small enough bending displacement, which is a constraint for practical applications. In our previous works [11, 18], a patterned IPMC strip was used to eliminate the BR effect. There were two regions on IPMC where the step voltage was applied to one region and the other's disturbance. There were two problems with the method. First, it is hard to apply two simultaneous voltages to small-size patterned electrode IPMCs, and second, by etching a pattern, we lose some part of IPMC, which causes less bending displacement for a constant step voltage. This paper presented a developed version of our suggested method in reducing the IPMC's back relaxation [11, 18]. This method is based on the rapid reciprocally moving free water molecules in the Nafion membrane using a relatively high-frequency disturbance. Instead of using a patterned IPMC, a non-patterned one has been used for this experiment.

The rest of this paper is consisted of three sections. Our idea is described in Section II, which is based on a physicochemical-mathematical approach. In Section III, the proof of our proposed method is discussed, and in Section IV, the discussion is provided, followed by the conclusion.

## II. PROPOSED METHOD

### A. Back Relaxation (BR)

Before we start the explanations, we need to know what would happen in IPMC if we apply a voltage to it. According to Figure 1 there are two stages in the bending procedure of an IPMC actuator. The first stage is called inflation. During this stage, the movement of hydrated sodium cations is toward the cathode electrode, which is due to the applied voltage to the membrane. Due to this movement, water molecules' concentration is increased in the membrane's cathode side. On the contrary, the concentration decreases on the anode side. This migration stops when the positive charge of sodium ions becomes equal to the cathode's negative charge. Because there are more water molecules at the cathode side than the anode side, the IPMC will be bent toward the anode, and the actuation will be observed [11]. During the inflation stage, the next stage will show up, called the back-relaxation stage. During this stage, hydrated sodium cations and water molecules want to move toward the cathode, respectively, because of the self-diffusion effect. This phenomenon intends to keep the concentration over the membrane at equilibrium, and consequently, back relaxation occurs [1, 19-21]. We should concentrate on this stage to find a solution to the back-relaxation phenomenon.

### B. Proposed BR elimination method

In the previous work [18], we presented a new method to eliminate the BR effect by using an IPMC with patterned electrodes and applying the desired voltage and the required disturbance to each electrode region. The idea was proposed that an exogenous disturbance should be applied to IPMC's membrane to create perturbation in water molecules. By theoretical explanation, the voltages that should be applied to each region had been calculated. Because of the problems introduced in the previous section, we want to use a non-pattern IPMC, and for this purpose, the theoretical explanation of the previous paper [18] should be rewritten.

From Figure 1 we assume that an electric field is applied to the IPMC. Therefore, x=0 represent the anode and x=h is the cathode. This electric field contains two individual electric fields: a step voltage and the other is the disturbance. As represented in Figure 2 the hydrated sodium cations, which can move freely inside the Nafion, are practically affected by three internal forces:

- Electric field force (Fe)
- Viscoelastic resistance force (Fv)
- The diffusion resistance force (Fd)

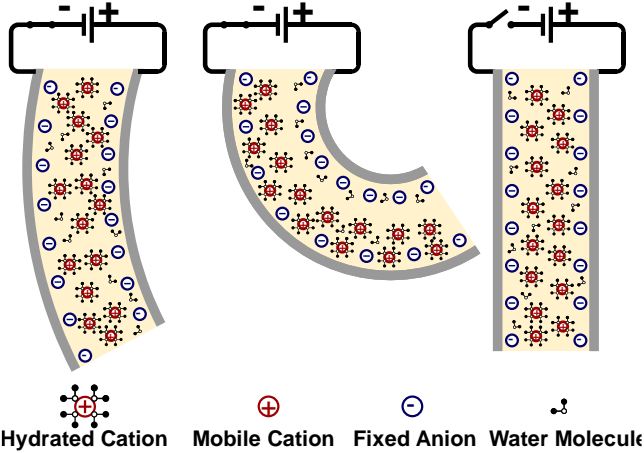

**Figure. 1. A 3D representation of an IPMC strip. (Right) before applying voltage, (Middle) the maximum bent IPMC after applied voltage, (Left) the bent IPMC after back relaxation occurred while the voltage is applying.**

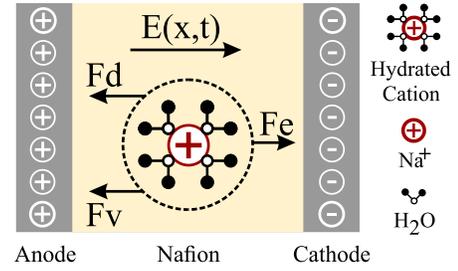

**Figure. 2. Representation of the internal forces in the IPMC membrane.**

From [18], we know that we can find the total charge of sodium ions from Eq. (1).

$$\frac{\partial Q(x,t)}{\partial t} = m^2 \frac{\partial Q^2(x,t)}{\partial x^2} + F(x,t) \quad (1)$$

Where $F(x,t)$ and $H(x,t)$ are defined by Eqs. (2) and (3).

$$F(x,t) = H(x,t)\frac{\partial Q(x,t)}{\partial x} \quad (2)$$

$$H(x,t) = m^2 \frac{\partial \ln(w(x,t))}{\partial x} \\ - \frac{e}{\varepsilon\eta}\left(\frac{1}{A_x}\int_0^t i(\tau)d\tau + Q(x,t) - Q(x,0)\right) \quad (3)$$

And

$$m = \sqrt{KT/\eta} \quad (4)$$

Where $Q(x,t)$ is the total charge of sodium ions, $w(x,t)$ is the concentration of water molecules, $e$ is the elementary charge, $\varepsilon$ is the dielectric constant of the Nafion membrane, $\eta$ is the viscosity constant of the hydrated sodium ions, $A_x$ is the cross-section area of the Pt layer, $i(t)$ is applied electric current, $K$ is the Boltzmann coefficient, and $T$ is the thermodynamic temperature.

To solve Eq. (1), we need to define initial and boundary conditions. These conditions define the value of the total charge of the sodium ions in both side of the anode and cathode and also at the beginning of the time, as shown in Eqs. (5) and (6).

$$Q(x,0) = NeA_x c_0 x \quad (5)$$

$$\begin{cases} Q(0,t) = 0 & x = 0 \quad t > 0 \\ Q(h,t) = NeA_x c_0 h & x = h \quad t > 0 \end{cases} \quad (6)$$

Where $c_0$ is the concentration of anions, which is constant because the anions are fixed in the backbone of Nafion, and so their concentrations don't change over time.

By solving Eq. (1), we can calculate the value of the concentration of sodium ions. From [18], we know that by considering the steady-state condition, Eq. (7) can be obtained from E1. (1).

$$c(x,t) = -\frac{1}{m^2}\int_0^x H(\psi,t)c(\psi,t)d\psi \quad (7)$$

By the Ohm's law we can specify the applied electric current:

$$i(t) = i_1(t) + i_2(t) \quad (8)$$

$$i_1(t) = G v_1(t) \quad (9)$$

$$i_2(t) = G v_2(t) \quad (10)$$

Where $i_1(t)$ is the current of the step input, $i_2(t)$ is the current of the disturbance, and G is the surface conductivity of IPMC.

By replacing $H(x,t)$ with Eq. (3) in Eq. (7) and taking derivation with respect to t and x and ignore small terms, just like the previous work, we reach Eq. (11).

$$\frac{\partial^2 c(x,t)}{\partial x \partial t}\left(1+\ln\left(\frac{w(x,t)}{w(0,t)}\right)\right)+\frac{\partial c(x,t)}{\partial t}\left(\frac{\partial}{\partial x}\ln\left(\frac{w(x,t)}{w(0,t)}\right)\right)$$
$$+\frac{\partial c(x,t)}{\partial x}\left(\frac{\partial}{\partial t}\ln\left(\frac{w(x,t)}{w(0,t)}\right)\right)+c(x,t)\left(\frac{\partial^2}{\partial x \partial t}\ln\left(\frac{w(x,t)}{w(0,t)}\right)\right) \quad (11)$$
$$+c(x,t)\left(\frac{eG}{\varepsilon KTA_x}(v_1(t)+v_2(t))\right)=0$$

As discussed, the hydrated sodium cations reciprocate between anode and cathode, so we need to specify the value of $v_2$. In order to create a uniform distribution of water molecules all over the membrane, we need to apply a specific voltage ($v_2$) which perturbed the free water molecules in the whole Nafion membrane. So, in the steady-state, the free water molecules concentration is homogenous in the Nafion:

$$\frac{w(x,t)}{w(0,t)} \approx 1. \quad (12)$$

Because $v_1(t)$ is a step voltage, we can solve Eq. (11) and reach the solution of the concentration, which is provided in Eq. (13) (Details have been described in Appendix A).

$$c(x,t) = \hat{k}\, e^{\alpha x - \Delta(t)} \quad (13)$$

Where

$$\Delta(t) = \frac{\Gamma G}{\alpha}\left(V_p t + \int v_2(t) dt\right) \quad (14)$$

$$\Gamma = \frac{e}{\varepsilon K T A_x} \quad (15)$$

$\hat{k}$ and $\alpha$ are constants based on the given initial and boundary conditions and $V_p$ is the amplitude of the step voltage ($v_1$). In order to get the steady-state concentration of hydrated sodium cations, we can use Eq. (13). We expect that the concentration should be constant near the cathode. To satisfy this expectation, the value of α should be positive.

$$\lim_{(x,t)\to(h,\infty)} c(x,t) = \hat{c} \quad (16)$$

$$\alpha > 0 \quad (17)$$

As discussed earlier, Nafion-based IPMCs suffer from the BR effect, which can be reduced by applying an external voltage ($v_2$) as a disturbance. Applying an external disturbance will not cause any instability to the ionic distribution throughout the membrane. The displacement dynamic of IPMC and the hydrated sodium cations' movement dynamic are not comparable, and the displacement dynamic is significantly slower than the movement dynamic. Despite the hydrated sodium cations' movement, the overall displacement is stationary concerning the IPMC's bending because ions in the membrane move much faster than the IPMC's bending speed. As mentioned before, the dynamic of hydrated cations should be swifter compared to the dynamic of IPMC itself to eliminate the BR effect. To investigate this occurrence, we need to find an equation that can describe molecules' velocity inside the membrane. By taking the first derivative of Eq. (13) with respect to t, we reach to Eq. (18).

$$v_c(x,t) = \frac{\partial c(x,t)}{\partial t} = \frac{-\Gamma G}{\alpha}(V_p + v_2)\hat{k} e^{\alpha x - \Delta(t)} \quad (18)$$

Now we have to talk about $v_2$. Let us consider that the amplitude of $v_2$ is approximately the same as the value of $V_p$. It is needed to make assumptions on the speed of water molecules. As seen from Eq. (18), we can define the value of water molecules' velocity. We can then assure that the value of the speed of perturbation of water molecules in the membrane is high enough to satisfy our assumptions. Also, suppose its frequency is relatively higher than the frequency of the IPMC's tip displacement. In that case, we can see that water molecules' movement and their perturbation can be considered working separately. In this situation, there would be a section at the middle of IPMC, where water molecules will steer continuously and make the perturbation that we want.

So, these assumptions about frequency and amplitude would give us a proper disturbance signal, which can be applied to our input to make the desired output.

## III. EXPERIMENTAL RESULTS

It is needed to evaluate our experimental data with some primary and advanced criteria to verify the proposed method. First, we introduce our hardware apparatus, and then the required criteria and their result on the data were presented.

### A. Hardware apparatus

In Figure 3 the experimental apparatus is shown. It contains several components which are listed:
- Computer with MATLAB 2020a
- Power supply
- Adjustable mechanical camera holder
- Camera
- IPMC holder
- IPMC
- Arduino microcontroller
- MCP4725 Digital to Analog
- Analog circuits

The desired signal is generated in MATLAB and sent to our analog circuit via Arduino Uno and MCP4725. The input of the analog circuit change by amplitude and offset to the desired values. When the signal is applied to the IPMC, a video will capture the IPMC's functioning by the camera. Then, we can extract data from these videos by processing techniques [11]. In Figure 4 which is also discussed in [22], the electronic circuit's schematic view is shown with its inputs. The place of IPMC in the background would capture by the camera, and the coordination of IPMC's tip stored for further processes.

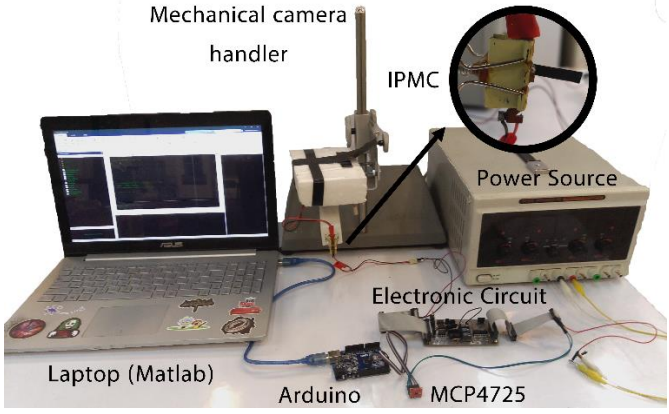

**Figure. 3. data acquisition setup.**

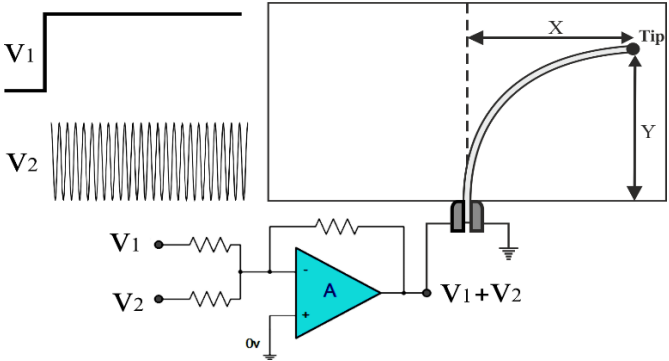

**Figure. 4. Schematic of extracting features from IPMC's response to the applied signal. V1 is the step voltage, and V2 is the disturbance. X and Y are the tip displacement in the x and y directions. [22]**

*B. Results*

The IPMC used in this experiment is made of Nafion 117, and its electrodes are Platinum. The size of the IPMC is 23×3.5×0.2 mm³. As discussed in [11], for observing the BR effect reduction in the IPMC, the step response should show more stability and contains less slope. To this end, a step voltage ($v_1 = 4V$) is produced and then applied to the IPMC in addition to a specified signal ($v_2$). We apply the disturbance as $v_2$ and observe the response to the step voltage for values of $v_2$, 0.25, 0.5, and 1 time of $v_1$ amplitude, and different frequencies of disturbance, including 60, 600, 1500, 3000, 4000, 6000 Hz.

As mentioned before, our suggested signal for $v_2$ is a relatively high-frequency disturbance ("High" compared to the working frequency of IPMC). This specific disturbance stirs the water molecules in the membrane during the IPMC bending process. The step response component along the X-axis has been shown in Figure 5, respectively, for all inputs. Also, the maximum displacement of IPMC's tip and final displacement of IPMC's tip have been displayed in Figure 7 and Figure 8.

The best performance criteria in the evaluation of the time domain step response are its components. These components are settling time ($t_{settle}$), rise time ($t_{rise}$), and percentage overshoot ($PO$). It is desired that all of these criteria should be minimized. Therefore, a geometric average of all the step response criteria is introduced as a single criterion to achieve an acceptable step response [23].

$$g_\mu = \sqrt[3]{PO \cdot t_{settle} \cdot t_{rise}} \quad (19)$$

Another evaluation of the step response goodness is fuzzy logic criteria [18]. Fuzzy logic converts all of the expert knowledge of a problem to mathematical propositions, which can be used in system modeling and managing uncertainty [24-27]. In recent works [11, 18], fuzzy logic criteria have been used to evaluate the IPMC's step response. Any single step response criteria cannot evaluate the step response quality because none is sufficient for evaluation, and also, the geometrical average ($g_\mu$) is not acceptable [11, 18]. The criterion propose here is the goodness of the step response ($\phi$), which is described as below:

$$\phi = f(t_{rise}, t_{settle}, PO) \quad (20)$$

Here, $f(.)$ is a nonlinear fuzzy mapping, realized by a Mamdani product inference engine [11]. The BR reduction amount can be explicated using the step response criteria ($t_{rise}$, $t_{settle}$, and $PO$). The step response is considered good if all three criteria are good, and then the BR effect is suppressed, but they do not affect the fuzzy criterion equally. On the contrary, the step response is considered not good if PO is mediocre, but $t_{rise}$ and $t_{settle}$ are in a reasonable range because the overshoot has more impact on the goodness of the step response ($\phi$) compared to other criteria. The fuzzy assessment used here described in [11], both itself and its membership functions, which are used here because that has shown promising results for [11, 18].

The step response performance criteria of the IPMC's functioning and the geometric average of the step response criteria have been presented in Figure. 9 to Figure. 1. Also, fuzzy criterion ($\phi$) shown in Figure. 13.

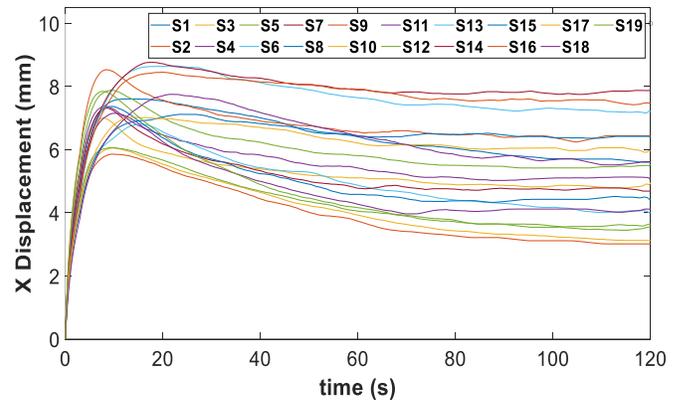

**Figure. 5. Tip displacement of the IPMC in X-direction in response to a step voltage of $v_1$ which is 4V. S1-6, $v_2$ amplitude is $0.25 * v_1$. S7-12, $v_2$ amplitude is $0.5 * v_1$. S13-18, $v_2$ amplitude is $v_1$. Frequency of each group is 60, 600, 1500, 3000, 4000, and 6000. S19 is data without disturbance.**

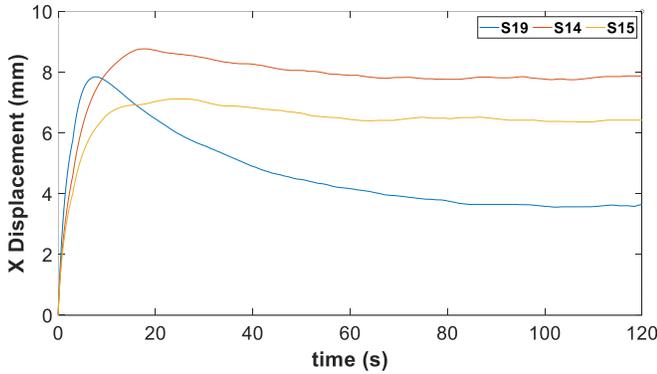

**Figure. 6.** Tip displacement of the IPMC in X-direction in response to a step voltage of $v_1$ which is 4V. In S14, $v_2 = v_1$ and frequency is 600Hz. In S15, $v_2 = v_1$ and frequency is 1500Hz. In S19, there is no $v_2$ and IPMC works conventionally.

To begin our comparison for choosing the best noisy input, we must start from the most straightforward criterion that we introduced, and that criterion is step response. It is needed to name the inputs to simplify the next part. Each S is input with the same step voltage of $v_2 = 4V$ with various disturbance amplitude and frequency values. The result of x displacement for each input is shown in Figure 5. The best three curves shown in Figure 5 are S13, S14, and S16, which have the least back relaxation effect observed by the figure with naked eye. If we look at their step response criteria (Figures 9, 10, and 11), the values for S14 are much more suitable than S13 and S16. Surprisingly, S15 also has outstanding results in terms of step response criteria. Remember that the rise time value (Figure 10) is high for the inputs mentioned because the amplitude of disturbance voltage will prevent the IPMC from bending fast. The curve of S15 in Figure5 has lower x displacement values than S14 but similar values of step response criteria. Their x displacement is traced in Figure 6 with the input without disturbance, which is S19. As it can be seen, all the criteria discussed earlier cannot evaluate the goodness of the step response alone. It is needed to specify the step response with another criterion that contains all of the step response components. Here, we use the geometric average of step response components [18], which is shown in Figure 12. For this one, the best value, the minimum value, belongs to both S14 and S15, which is not helpful at all. Hence, the fuzzy logic assessment has been used, which is described earlier. By required assumptions, the value of the fuzzy logic criterion is 90 for both S14 and S15. Fuzzy logic shows us that the goodness of S14 and S15 is more than other inputs, which can be perceived previously. We describe new criteria to evaluate the best result for us by maximum displacement and final displacement of IPMC's tip, which are tip displacement's essential values shown in Figures 7 and 8. First, the percentage of displacement disputes from the maximum value to the final value should be checked. It can be calculated from Eq. (56):

$$PD = \frac{MD - FD}{FD} \quad (21)$$

Where "MD" is the maximum displacement of IPMC and "FD" is the final displacement of IPMC. The value of maximum displacements is shown in Figure 7, and the final displacements in Figure 8. By calculating PD for S14 and S15, it can be demonstrated that both values are 0.11. In this situation, when none of our criteria can decide between S14 and S15, we can concentrate just on the maximum displacement of IPMC. This parameter is crucial for us. First, it is essential because we don't want too much displacement drop because of adding disturbance to step voltage; second, we pursue eliminating the BR effect in large displacement. By comparing the maximum displacement of S14 and S15, S14 has 23 percent more displacement than S15. Also, this value for the final displacement is 23 percent. This result can also be seen in Figure 6. So overall, S14 ($v_2 = v_1$ and frequency is 600 Hz) has the best outcome in the elimination of the BR effect.

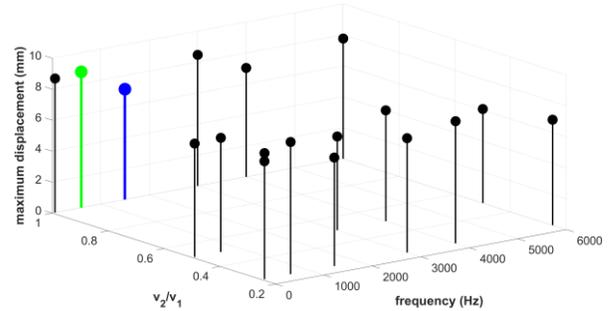

**Figure. 7.** Maximum displacement of IPMC. Green data is $v_2 = v_1$ and $f = 600\ Hz$ and its value is 8.7678mm. Blue data is $v_2 = v_1$ and $f = 1500\ Hz$ and its value is 7.1209mm. Maximum value of this chart belongs to green data.

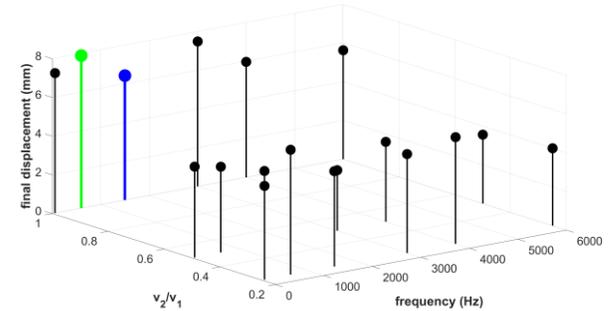

**Figure. 8.** Final displacement of IPMC. Green data is $v_2 = v_1$ and $f = 600\ Hz$ and its value is 7.8794mm. Blue data is $v_2 = v_1$ and $f = 1500\ Hz$ and its value is 6.4223mm. Maximum value of this chart belongs to green data.

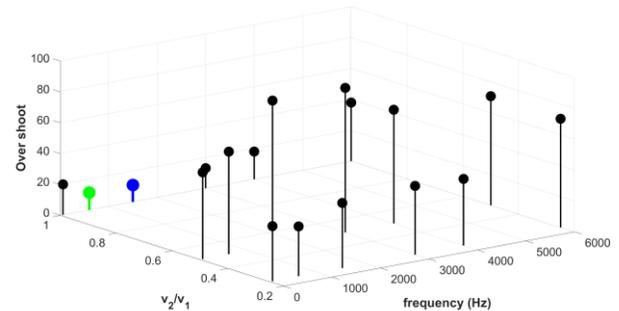

**Figure. 9.** Overshoot of IPMC. Green data is $v_2 = v_1$ and $f = 600\ Hz$ and its value is 11.2751. Blue data is $v_2 = v_1$ and $f = 1500\ Hz$ and its value is 10.8778. Minimum value of this chart belongs to both green and Blue data.

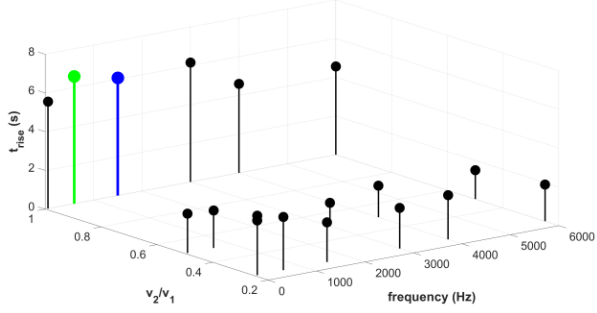

Figure. 10. Rise time of IPMC. Green data is $v_2 = v_1$ and $f = 600\ Hz$ and its value is 6.5710s. Blue data is $v_2 = v_1$ and $f = 1500\ Hz$ and its value is 6.0745s. Minimum value of this chart belongs to neither blue data or green data and its value is 1.2098s.

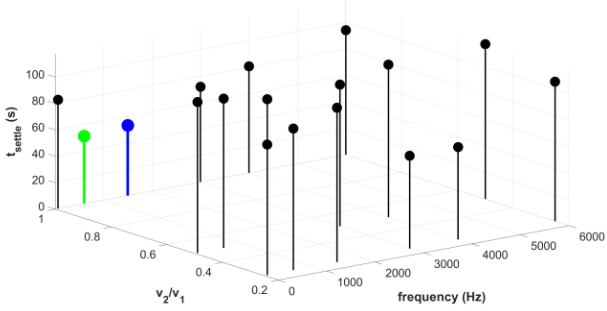

Figure. 11. Settling time of IPMC. Green data is $v_2 = v_1$ and $f = 600\ Hz$ and its value is 51.1113s. Blue data is $v_2 = v_1$ and $f = 1500\ Hz$ and its value is 53.1050s. Minimum value of this chart belongs to green data.

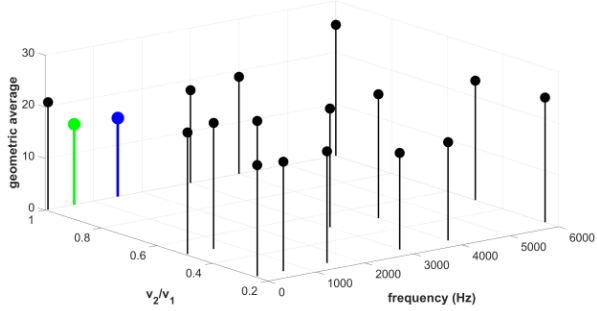

Figure. 12. Geometric average of step response criterion of IPMC. Green data is $v_2 = v_1$ and $f = 600\ Hz$ and its value is 15.4868. Blue data is $v_2 = v_1$ and $f = 1500\ Hz$ and its value is 15.1960. Minimum value of this chart belongs to both green and blue data.

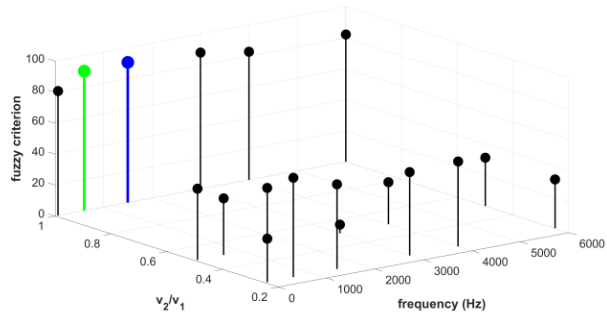

Figure. 13. Fuzzy criterion of IPMC. Green data is $v_2 = v_1$ and $f = 600\ Hz$ and its value is 90.1. Blue data is $v_2 = v_1$ and $f = 1500\ Hz$ and its value is 90.5. Maximum value of this chart belongs to both green and blue data.

## IV. CONCLUSION

A method for eliminating the unwanted back relaxation effect of pattern-free IPMC was presented, and by an analytical and experimental investigation, the validity of the method was proved. We experimentally and theoretically proved that, like the patterned IPMCs, by applying a relatively high-frequency disturbance to the pattern-free IPMCs, the hydrated sodium cations will reciprocate rapidly between cathode and anode electrodes result in a perturbation of the free water molecules in the IPMC membrane. This fast reciprocation holds free hydrated cations at the cathode side and helps to reduce the BR effect tremendously in Nafion-based IPMCs. The results showed that the proposed BR elimination approach could significantly reduce the BR effect of IPMC, needless to have any pattern on the electrodes of the IPMCs, and unlike the patterned IPMCs, no attenuation will happen in the tip displacement of IPMC, and even we saw the increased tip displacement.

## V. APPENDIXES

### A. An analytical solution for reaching Eq. (13):

For reaching Eq. (13) using Eq. (11), we need to use the physical knowledge we have from IPMC. A fast-reciprocating movement of hydrated sodium cation will be created in the membrane by applying a disturbance with the primary signal. In this situation, the concentration of water molecules is almost uniform in the membrane because they are stirred rapidly by the disturbance. Consequently, the free water molecules concentration is almost identical in the steady-state in the membrane of IPMC. As a result, it can be inferred that the free water molecules concentration is uniform throughout the membrane ($w(x,t)$) including the area around the Anode ($w(0,t)$), which can be describe as:

$$\frac{w(x,t)}{w(0,t)} \approx 1 \tag{a1}$$

And respectively:

$$\ln\left(\frac{w(x,t)}{w(0,t)}\right) \approx 0 \tag{a2}$$

$$\frac{\partial}{\partial t}\ln\left(\frac{w(x,t)}{w(0,t)}\right) \approx 0 \tag{a3}$$

$$\frac{\partial}{\partial x}\ln\left(\frac{w(x,t)}{w(0,t)}\right) \approx 0 \tag{a4}$$

By utilizing equations (a1) to (a4) into Eq. (11) we can obtain following equation:

$$\frac{\partial^2 c(x,t)}{\partial x \partial t} + \left(\frac{eG}{\varepsilon KTA_x}\left(v_1(t) + v_2(t)\right)\right)c(x,t) = 0 \tag{a5}$$

Finally by using the separation of variables method, equation (a5) can be solved into Eq. (13).

## REFERENCES

[1]    P.Brunetto, L.Fortuna, P.Giannone, S.Graziani, F.Pagano, "A resonant vibrating tactile probe for biomedical applications based on

IPMC," *IEEE Transactions on Instrumentation and Measurement,* vol. 59, No. 5, pp. 1453-1462, 2010.

[2] G.H.Feng, and J.W.Tsai, "Micromachined optical fiber enclosed 4-electrode IPMC actuator with multidirectional control ability for biomedical application," *Biomed Microdevices,* pp. 169-177, 2011.

[3] Y.-c. Chang and W.-j. Kim, "Aquatic Ionic-Polymer-Metal-Composite Insectile Robot With Multi-DOF Legs," *IEEE/ASME Transactions On Mechatronics,* vol. 18, no. 2, pp. 547-555, 2013.

[4] S. Ruiz, B. Mead, V. Palmre, K. J. Kim, and W. Yim, "A cylindrical ionic polymer-metal compositebased robotic catheter platform: modeling, design and control," *Smart Materials and Structures,* vol. 24, 2015.

[5] S. Ford, G. Macias, and R. Lumia, "Single active finger IPMC microgripper," *Smart Materials and Structures,* vol. 24, 2015.

[6] R.K. Jain, S.Majumder and A.Dutta, "Micro assembly by an IPMC based flexible 4-bar mechanism," *Smart Materials and Structures,* vol. 21, No. 7, 2012.

[7] J. J. Hubbard, M. Fleming, K. K. Leang, V. Palmre, D. Pugal, and K. J. Kim, "Characterization Of Sectored-Electrode Ipmc-Based Propulsors For Underwater Locomotion," in *ASME 2011 Conference on Smart Materials, Adaptive Structures and Intelligent Systems*, Scottsdale, Arizona, USA, 2011.

[8] M. Annabestani, P. Esmaeili-Dokht, S. K. Nejad, and M. Fardmanesh, "NAFAS: Non-rigid Air Flow Active Sensor, a cost-effective, wearable, and ubiquitous respiratory bio-sensor," *IEEE Sensors Journal,* 2021.

[9] S.Tadokoro , S.Yamagami and T.Takamori, "An actuator model of ICPF for robotic applications on the basis of physico-chemical hypotheses," *Proceeding of IEEE International Conference on Robotics and Automation-San Francisco-USA,* pp. 1340–1346, 2000.

[10] Y.Gong, J.Fan, C.Y. Tang, and C.P.Tsui, "Numerical Simulation of Dynamic Electro-Mechanical Response of Ionic Polymer-Metal Composites," *Journal of Bionic Engineering,* pp. 263–272, 2011.

[11] M. Annabestani, M. Maymandi-Nejad, and N. Naghavi, "Restraining IPMC Back Relaxation in Large Bending Displacements: Applying Non-Feedback Local Gaussian Disturbance by Patterned Electrodes," *IEEE Transactions on Electron Devices,* vol. 63, no. 4, pp. 1689-1695, 2016, doi: 10.1109/TED.2016.2530144.

[12] S.Nemat-Nasser, "Micromechanics of actuation of ionic polymer-metal composites," *Journal of Applied physics,* vol. 92, No. 5, pp. 2899-2915, 2002.

[13] S.-M. Kim and K. J. Kim, "Palladium buffer-layered high performance ionic polymer–metal composites," *Smart Materials And Structures,* vol. 17, 2008.

[14] S. Nemat-Nasser and S. Zamani, "Effect of solvents on the chemical and physical properties of ionic polymer-metal composites," *Journal Of Applied Physics* vol. 99, 2006.

[15] D. B. Nikhil and K. Won-jong, "Precision Position Control of Ionic Polymer Metal Composite," in *American Control Conference* Boston, Massachusetts, 2004, pp. 740-745.

[16] K. Takagia, S. Hirayama, S. Sano, N. Uchiyama, and K. Asaka, "Reduction of the stress-relaxation of IPMC actuators by a fluctuating input and with a cooperative control," in *Electroactive Polymer Actuators and Devices (EAPAD)*, Y. Bar-Cohen, Ed., 2012, doi: 10.1117/12.915161.

[17] M. J. Fleming, K. J. Kim, and K. K. Leang, "Mitigating IPMC back relaxation through feedforward and feedback control of patterned electrodes," *Smart Materials And Structures,* vol. 21, 2012, doi: 10.1088/0964-1726/21/8/085002.

[18] M. Annabestani, N. Naghavi, and M. Maymandi-Nejad, "From modeling to implementation of a method for restraining back relaxation in ionic polymer–metal composite soft actuators," *Journal of Intelligent Material Systems and Structures,* vol. 29, p. 1045389X1878308, 07/24 2018, doi: 10.1177/1045389X18783082.

[19] C.Bonomo, P.Brunetto, F.Fortuna, P.Giannone, S.Graziani and S.Strazzeri, , "A tactile sensor for biomedical applications based on IPMCs," *IEEE Sensors Journal,* vol. 8, No. 8, pp. 1486-1493, 2008.

[20] M.Shahinpoor, P.Shahinpoor and D.Soltanpour, "Surgical Correction Of Human Eye Refractive Errors By Active Composite Artificial Muscle Implants," *United States Patent/ US 6511508 B1,* 2003.

[21] Y.Bahramzadeh, and M.Shahinpoor, "Ionic polymer-metal composites (IPMCs) as dexterous manipulators and tactile sensors for minimally invasive robotic surgery," in *Proceeding of SPIE 19th Annual International Symposium on Smart Structures and Materials, San Diego, California*, 11-15 March, 2012.

[22] M. Annabestani, M. H. Sayad, P. Esmaeili-Dokht, R. Gorji, and M. Fardmanesh, "Eliminating Back Relaxation in Large-Deformable IPMC Artificial Muscles: A Noise-Assistive Pattern-Free Electrode Approach," in *2020 27th National and 5th International Iranian Conference on Biomedical Engineering (ICBME)*, 26-27 Nov. 2020 2020, pp. 55-60, doi: 10.1109/ICBME51989.2020.9319432.

[23] F.Carpi , and D.De Rossi, "Electroactive Polymer-Based Devices for e-Textiles in Biomedicine," *IEEE Transactions On Information Technology In Biomedicine,* vol. 9,no 3, 2005.

[24] M. Annabestani, A. Rowhanimanesh, A. Mizani, and A. Rezaei, "Fuzzy descriptive evaluation system: real, complete and fair evaluation of students," *Soft Computing,* vol. 24, no. 4, pp. 3025-3035, 2020/02/01 2020, doi: 10.1007/s00500-019-04078-0.

[25] M. Annabestani and M. Saadatmand-Tarzjan, "A New Threshold Selection Method Based on Fuzzy Expert Systems for Separating Text from the Background of Document Images," *Iranian Journal of Science and Technology, Transactions of Electrical Engineering,* vol. 43, no. 1, pp. 219-231, 2019/07/01 2019, doi: 10.1007/s40998-018-0160-7.

[26] F. Hasanzadeh and F. Faradji, "An ICA Algorithm Based on a Fuzzy Non-Gaussianity Measure," in *1st International Conference on New Research Achievements in Electrical and Computer Engineering*, 2016.

[27] F. Hasanzadeh, M. Annabestani, and S. Moghimi, "Continuous emotion recognition during music listening using EEG signals: A fuzzy parallel cascades model," *Applied Soft Computing,* vol. 101, p. 107028, 2021/03/01/ 2021, doi: https://doi.org/10.1016/j.asoc.2020.107028.

vol. 65, no. 12, pp. 2866-2872, 2016.